\documentclass[final,authoryear,12pt]{elsarticle}

\usepackage{graphicx}
\usepackage{color,soul}
\usepackage{epstopdf}
\usepackage{relsize}
\usepackage{booktabs}
\usepackage{natbib} 
\sethlcolor{green}
\usepackage{amssymb}
\usepackage{amsthm}
\usepackage{amsmath}
\usepackage{float}

\begin{document}
\begin{frontmatter}

\title{Contagion among Central and Eastern European stock markets during the financial crisis}

\author[ies,utia]{Jozef Barun\'ik} \ead{barunik@utia.cas.cz}
\author[utia,ies]{Luk\'a\v{s} V\'acha\corref{cor2}} \ead{vachal@utia.cas.cz}

\cortext[cor2]{Corresponding author} 
\address[utia]{Institute of Information Theory and Automation, Academy of Sciences of the Czech Republic, Pod Vodarenskou Vezi 4, 182 00, Prague, Czech Republic} 
\address[ies]{Institute of Economic Studies, Charles University, Opletalova 21, 110 00, Prague,  Czech Republic}

\begin{abstract}
This paper contributes to the literature on international stock market comovements and contagion. The novelty of our approach lies in application of wavelet tools to high-frequency financial market data, which allows us to understand the relationship between stock markets in a time-frequency domain. While major part of economic time series analysis is done in time or frequency domain separately, wavelet analysis combines these two fundamental approaches. Wavelet techniques uncover interesting dynamics of correlations between the Central and Eastern European (CEE) stock markets and the German DAX at various investment horizons. The results indicate that connection of the CEE markets to the leading market of the region is significantly lower at higher frequencies in comparison to the lower frequencies. Contrary to previous literature, we document significantly lower contagion between the CEE markets and the German DAX after the large 2008 stock market crash.
\end{abstract}

\end{frontmatter}

\section{Introduction}
International stock markets are witnessing increasing interconnection. As the stock markets are becoming more open to foreign investors, the increase in liquidity and availability of stocks of transition and emerging countries enlarges the possibilities of an international portfolio diversification. On the other hand, the integration and comovements are becoming stronger in time reducing these possibilities. The events between years 2007 and 2009 reminded us about possible reducing potential of the diversification during crisis, as interconnection of markets increased in this period.

Focusing on the time dimension of the market dynamics, researchers often ignore the dynamics at different investment horizons. These may be especially important, as they represent trading frequencies of investors with heterogenous beliefs. Starting with noise traders with an investment horizon of several minutes or hours, the spectrum of investors ranges through technicians with the horizon of several days to fundamentalists with the horizon of several weeks or months to investment funds with the investment horizon of several years. Thus, apart from the time domain, it is important to understand a frequency domain, which represents various investment horizons. As both domains are equally important and valid for deeper understanding of the financial markets' dynamics, one should not overlook them in the analysis. The importance of modeling the events in both domains motivates us to apply wavelet analysis which can work with both domains simultaneously.

In our work, we combine both time and frequency domain and we apply cross-wavelet analysis 
to study comovement and contagion on high-frequency (5-minutes) data. We concentrate on the Czech (PX), Hungarian (BUX) and Polish (WIG) stock indices with the benchmark German stock index (DAX), while we are interested mainly on the crisis period. The time frame chosen for the analysis give us an opportunity to study reaction of Central and Eastern European (CEE) financial markets to the large crash of September 2008. 

The literature on comovements, interdependence, and contagion is broad. The largest part of the literature examines interdependence between the US and countries of the Western Europe. \cite{Baele2005} and \cite{Baele2010} apply switching models to show that the intensity of comovements and spillovers increased during 1980s and 1990s with no evidence of significant contagion other than small effect during the 1987 crash. \cite{Connolly2007} research on comovements between the US, UK and German stock and bond markets and show that during high (low) implied volatility periods, the comovements are stronger (weaker) whereas stock-bond comovements tend to be positive (negative) following low (high) implied volatility days.  \cite{Morana2008} examine stock markets of the US, UK, Germany and Japan between 1973 and 2004 and finds increasing comovements for all markets.

The research of Central and Eastern European region is discussed in several works. \cite{Egert2007} examine high-frequency stock market comovements of the Czech, Hungarian, Polish, German, French and UK stock markets between 2003 and 2005. They find that correlations are much lower on high frequency data than daily data.  \cite{Gilmore2008} study the CEE stock markets and find strong cointegration but argue that signs of convergence to the Western Europe are lacking after the EU accession. \cite{gjika2013stock} examine the correlations between the Czech stock market and STOXX50 index on daily data, and they find that the correlation increased during the recent financial crisis. \cite{Hanousek2009} analyze high-frequency data of the CEE stock markets and show that these are strongly influenced by developed economies.

Whereas correlations and comovements are well defined through linkages based on fundamentals, definition of contagion varies across literature. \cite{reinhart1996capital} define ``fundamental--based" contagion as correlation over and above what one would expect from economic fundamentals whereas ``pure" contagion describes the transmission of shocks among countries in excess of what should be accounted to fundamental factors, i.e. it is characterized by excessive comovements \citep{bae2003new}. This type of contagion is usually caused by loss of confidence and panic in financial markets after an arrival of important negative news. Forbes and Rigobon (2002) define contagion in a similar way as a significant increase in cross-market linkages after a shock.

In our analysis of contagion we use approach of \cite{gallegati2012wavelet} that identifies contagion as a change in correlation structure between two time periods defined by large crash in September 2008. Using the wavelet correlation we decompose the correlation into various investment horizons. Recently, there are numerous works that research on contagion and comovement across financial markets using wavelets \citep{Conraria2008,RuaNunes2009,VachaBar2012}. 
\cite{ranta2013contagion} uses rolling wavelet correlation to uncover contagion among the major world markets during the last 25 years. Similar methodology was applied by \cite{dajcman2012comovement} to data of the CEE countries. Contagion between oil and stock markets in Europe and the US was studied by \cite{reboredo2013wavelet}, their results indicates contagion between oil and stock prices in Europe and the US since the onset of the global financial crisis. 

In our work, we enrich the current literature by employing the high frequency data to see whether there is strong interconnection of the studied markets also on higher frequencies that represents short trading horizons. In this respect, we follow our previous work studying the time-frequency correlations between gold, oil and stocks \citep{barunik2013gold}.

The main findings of this paper are that interconnection between all stock markets changed considerably in time and varies across investment horizons. We confirm the contagion between the DAX and PX, but we find unexpectedly lower correlations on high frequency investment horizons after the 2008 crash. This finding complements the literature studying the daily data, as most of the studies find increased correlations during the crisis period \citep{gjika2013stock}. Another interesting result of the wavelet correlation analysis is that the CEE markets generally exhibit low correlations on high frequencies, when compared to daily data. This shows that the CEE markets are connected to the leading markets in region only in terms of longer investment horizons.

The paper is structured as follows. In the next section we present very brief introduction to the  wavelet and contagion analysis. After the methodology is set, we describe the data. Next, we employ high-frequency data of the Czech (PX), Hungarian (BUX), Polish (WIG) and German (DAX) stock indices and study their comovement and contagion in the time--frequency domain. Final section concludes.

\section{Methodology}
This section briefly discusses wavelet techniques that are essential for the comovement and contagion analysis. Subsequently, we introduce formally the concept of contagion.

The wavelet transform allow us to decompose time series from the time domain to the time-frequency domain. Thus one dimensional time series are transformed into a two-dimensional space using localized function with a finite support, called wavelet. It is convenient for the decomposition in the situation when the time series under study is non-stationary, or only locally stationary \citep{Roueff2011}.  Important feature of wavelet analysis is the decomposition of the time series into frequency components called scales. With this decomposition we have an opportunity to study economic relationships on a scale-by-scale level which gives us a broader picture. Hence wavelets can be perceived as kind of ``lens" we have when studying the data. 

Particularly, we are able to separate the short-term and long-term investment horizons using wavelets. In the bivariate case, we are interested in the short-term and long-term comovements and dependencies. As the wavelets are localized functions we can also study dynamics of these relations in time. 

\subsection{Wavelet coherence}

As we study the comovement between two time series, we introduce a bivariate wavelet technique called wavelet coherence. First, we define the cross wavelet power of two time series $x(t)$ and $y(t)$ as 
\begin{equation}
|W_{xy} (u,j)| = W_x (u,s) \overline{W_y (u,j)},
\end{equation}
where $W_x (u,s)$ and $W_y (u,s)$ denote the continuous wavelet transforms of  time series $x(t)$ and $y(t)$, respectively, the bar denotes complex conjugate, parameter $u$ allocates a time position whereas parameter $j$ denotes the scale parameter. A low wavelet scale identifies the high frequency part of the time series -- short investment horizon. For example, the first scale, $j=1$ carries information about 10 minute investment horizons when using data with 5 minute frequency. For more details about wavelet transforms see \cite{Daubechies1988}, \cite{Mallat98} and \cite{PercivalWalden2000}. For the analysis of financial markets comovement, we use the Morlet wavelet.

The cross wavelet power uncovers areas in time-frequency space where the time series show a high common power. However, in the comovement analysis we search for areas where the two time series in time-frequency space comove, but does not necessarily have high power. Useful wavelet technique that find these comovements is the wavelet coherence.

Following \cite{TorrenceWebster99}, we define the squared wavelet coherence coefficient as
\begin{equation}
\label{R}
R^2 (u,j)=\frac{|S(j^{-1}W_{xy} (u,j))|^2}{S(j^{-1}|W_x (u,j)|^2) S(j^{-1}|W_y (u,j)|^2)},
\end{equation}
where $S$ is a smoothing operator\footnote{Smoothing is achieved by convolution in both time and scale, see \cite{Grinsted2004} for more details.}.
The coefficient $R^2 (u,j)$ is in the interval $[0,1]$. Values of the wavelet coherence close to one indicate strong correlation (denoted by red color in figures), while values close to zero (blue color in figures) indicate low or no correlation, see Figure \ref{fig:WTC HF}. The areas where the wavelet coherence is significant are bordered with the black thick contour\footnote{The theoretical distribution for the wavelet coherence is not known, the statistical significance is tested using Monte Carlo methods. The testing procedure is based on the approach of \cite{TorenceCompo98}.  In our analysis, we use the 5\% significance levels. }. The squared wavelet coherence coefficient can be perceived as a local linear correlation measure between two time series in the time--frequency space. 

As the wavelets are in fact filters, at the beginning and at the end of a dataset the filter analyzes nonexistent data. To solve this, we augment the dataset with a sufficient number of zeros. The affected area is called the cone of influence and it is graphically represented by a lighter-color area below the bold black line in figures. For more details, see \cite{TorenceCompo98}, \cite{Grinsted2004}.

The squared wavelet coherence coefficient can have positive values only, hence we cannot distinguish between negative and positive correlation directly. This problem can be solved using the wavelet coherence phase differences which indicate delays in the oscillation between the two time series, therefore we obtain information whether the two time series move together in phase (zero phase difference) or whether the time series are in anti-phase, i.e, they are negatively correlated. \cite{TorrenceWebster99} defines the wavelet coherence phase difference as
\begin{equation}
\phi_{xy} (u,j)=\tan ^{-1}\left( \frac{\Im \{S(s^{-1}W_{xy} (u,j))\}}{\Re \{S(s^{-1}W_{xy} (u,j))\} }\right),
\end{equation}
where $\Im$ and $\Re$ denote an imaginary and a real part operator, respectively. Phase differences are indicated by black arrows in the wavelet coherence figures in the areas with significant coherence.\footnote{Note that phase differences are depicted only for time-frequency areas with significant wavelet coherence.} In case the two examined time series move together on a particular scale and the arrows point to the right. If the time series are negatively correlated, then the arrows point to the left. Arrows pointing down (up) indicate that the first (second) time series leads the second (first) one by $\frac{\pi}{2}$.

\subsection{Wavelet correlation}

The wavelet correlation is computed using a discrete type of wavelet transform\footnote{We use the MODWT that is not restricted to sample sizes that are powers of two. For more details about the MODWT see \cite{PercivalWalden2000} and \cite{Gencay2002}}. Unlike the continuous wavelet transform, the discrete version of wavelet transform decomposes time series to vectors of wavelet coefficients that represents frequency bands. For example, wavelet coefficients at scale $j$ represent frequency band $f\in[1/2^{j+1},1/2^{j}]$, thus the highest frequency, the shortest investment horizon, characterize the first scale, $j=1$. Having vectors of discrete wavelet coefficients for all scales, $w(u,j)$ we can define wavelet correlation at scale $j$ for time series $x(t)$ and $y(t)$ as
\begin{equation}
\label{wvar}
\rho_{j} =\frac {Cov[w_x(u,j),w_y(u,j)]}  {Var[w_x(u,j)]Var[w_y(u,j)]}.
\end{equation}
The wavelet correlation provide a measure between two time series on scale-by-scale basis. 
In our study we use estimator of wavelet correlation based on Eq.(\ref{wvar}). For more detailed treatment of wavelet correlation and computation of confidence intervals, see \cite{Gencay2002}.

\subsection{Analysis of contagion}

Following \cite{forbes2002no} and \cite{gallegati2012wavelet} we define contagion as a change in correlation structure in two non--overlapping time periods. As a measure of correlation the wavelet multiresolution correlation is used. Formally, we define a null hypothesis of no contagion at a scale $j$ as
\begin{equation}
H_0: \rho_{(I),j} = \rho_{(II),j} \hspace{5mm} j=1,\ldots,J
\end{equation}
where $\rho_{(.),j}$ denotes wavelet correlation at scale $j$, the index $I$ and $II$ indicates the time period used for the correlation estimation. Since we decompose the 5 minute time series of stock market returns to eight wavelet scales, $J=8$, we can study contagion at investment horizons ranging from 10 minutes to 3 days.

\section{Data and empirical results}
\subsection{Data description}
In our analysis, we use 5-minute high-frequency data of the Czech (PX), Hungarian (BUX) and Polish (WIG) stock indices with a benchmark of German stock index (DAX). Central European stock markets data were collected over a period of 2 years beginning with January 2, 2008 and ending by November 30, 2009. The data were obtained from TICK data.

When looking at the data, one quickly observes that number of observations for each trading day differs among the indices. This problem arises due to different stock market opening hours. Prague Stock Exchange, as well as Warsaw Stock Exchange, is open from 9:30 to 16:00 Central European Time (CET). Budapest Stock Exchange is open from 9:00 to 16:30 CET. Finally, Frankfurt Stock Exchange is open from 9:00 to 17:30 CET. Thus we need to adjust the dataset by including only the periods of day where the data is available for all analyzed stock indices. We compute logarithmic returns for the period from 9:30 to 16:00 CET for each day separately in order to avoid overnight returns. Finally, we are left with 77 return observations for each stock market for each day of the analyzed period. By discarding major public holidays, the final sample includes 450 trading days.

\begin{table}[b]
\footnotesize
\begin{center}
\begin{tabular}{@{} lcccc @{}}
\toprule
  & DAX & PX & BUX & WIG \\
\midrule
 Mean & -2.98577$\times 10^{-6}$ & -1.66586$\times 10^{-5}$ & -2.63242$\times 10^{-5}$ & -1.76341$\times 10^{-5}$ \\
 St.dev & 0.00152547 & 0.0012775 & 0.00169863 & 0.0018456 \\
 Skewness & 0.426297 & -0.144049 & 0.131315 & 0.208045 \\
 Kurtosis & 20.3099 & 16.3973 & 27.0218 & 12.824 \\
 Min & -0.0156048 & -0.0146523 & -0.0243584 & -0.0177151 \\
 Max & 0.0317693 & 0.0170459 & 0.0434668 & 0.0276198 \\
 \bottomrule
\end{tabular}
\caption{Descriptive Statistics for 5-min high-frequency data.}
\label{stats}
\end{center}
\end{table}

\begin{figure}
\centering
\includegraphics[scale=0.53]{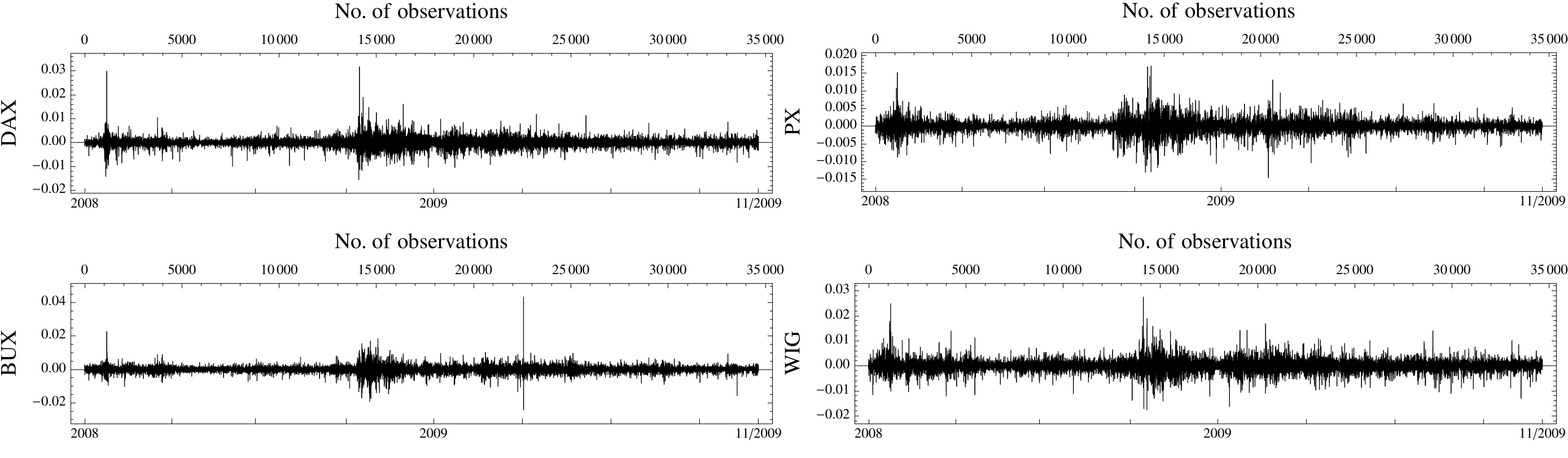}
\caption{Plots of 5-min logarithmic returns for DAX, PX, BUX and WIG indices.}
\label{ret}
\end{figure}

Table \ref{stats} provides descriptive statistics for our final sample of 5-minute high-frequency returns. Figure \ref{ret} shows plots of the data. 
\\

\begin{figure}
\centering
\includegraphics[scale=0.33]{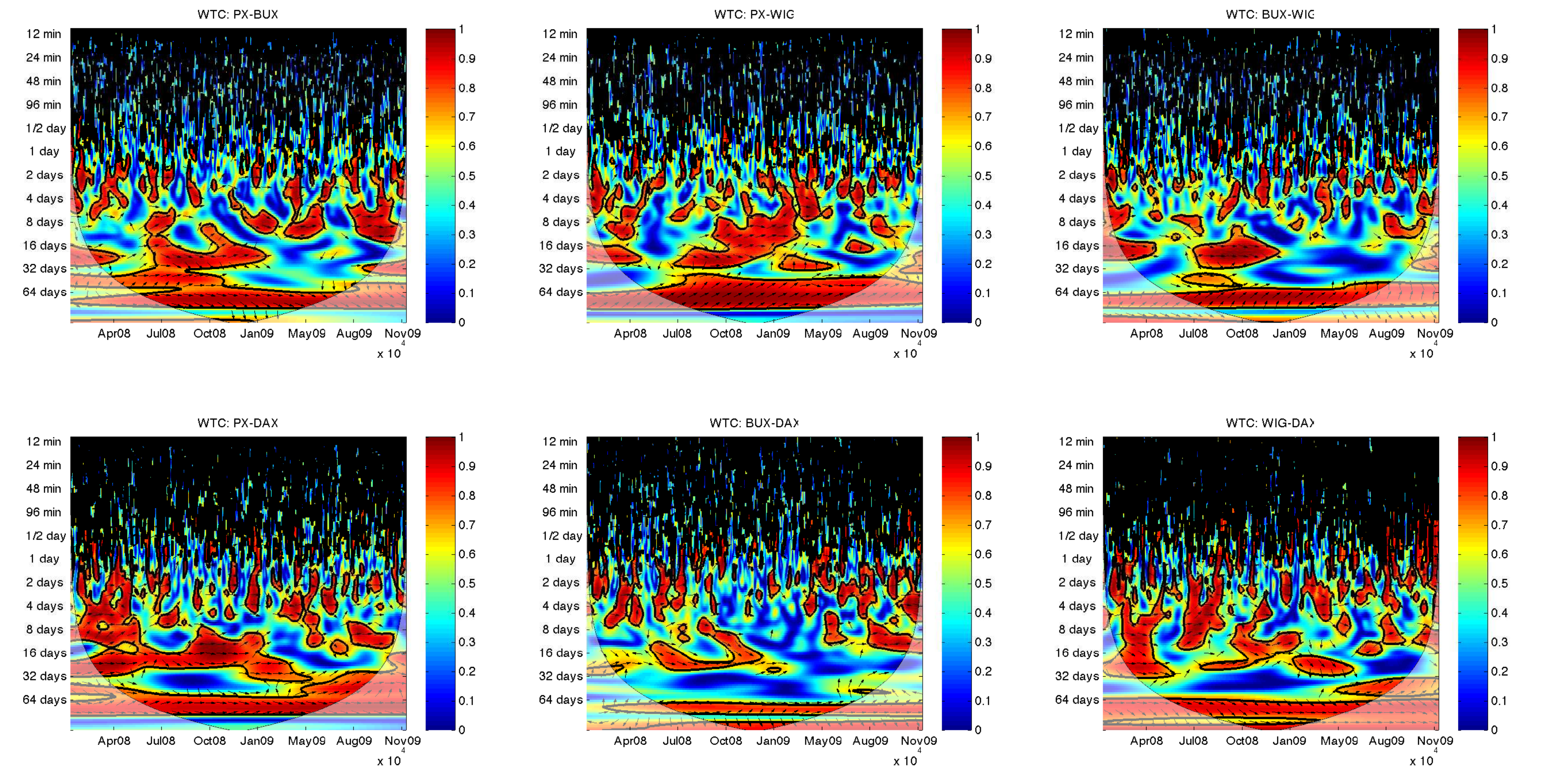}
\caption{Wavelet coherence of PX, BUX, WIG and DAX indices pairs on the 5 minutes high-frequency returns. Horizontal axis shows time, while vertical axis shows period in minutes/days. The warmer the color of region, the higher the degree of dependence between the pair.}
\label{fig:WTC HF}
\end{figure}

\subsection{CEE stock market comovements during the crisis}
Figure \ref{fig:WTC HF} shows the estimated wavelet coherence and the phase difference for all examined pairs of indices. Time is on the horizontal axis while vertical axis refers to frequency (the lower the frequency, the higher the scale, or period). Regions inside the black lines plotted in warmer colors represent regions where significant dependence has been found. The colder the color is the less dependent the series are. Blue regions represent periods and frequencies with no dependence in the indices. Thus the plot clearly identifies both frequency bands and time intervals where the series are highly coherent. The continuous wavelet transform at a given point uses  information of neighbor data points, thus areas at the beginning and at the end of the time interval should be interpreted with caution as we have discussed in the methodology part. Especially the time-frequency blocks inside the cone of influence on lower frequencies where the transform does not have sufficient number of data.

From the analysis of the wavelet coherence, we can observe interesting results. First of all, there are large significant comovement periods among all tested stock markets through several frequencies. As for the high-frequency patterns, it is hard to see from the figures as the black regions consist of many small periods of significant comovement at various frequencies (10min, 20min, etc.). Each of the pairs also shows strong comovement periods on several daily frequencies up to two and three weeks as well as periods where pairs comove on the several months scales.

When we look at the comovement of the PX, BUX and WIG (Figure \ref{fig:WTC HF}), we can observe that the PX is positively correlated with the WIG on lower frequencies up to several months. The PX-WIG pair also shows very interesting development of changing cross-correlation from the second half of year 2008 until the end of the first half of 2009. Correlations are strongly significant through this time period but they change from the month period (lower frequency) to the shorter period of one week (higher frequency). This dynamics of interdependence visible from wavelet transform of high-frequency data is unique and allows to understand the relationship between the analyzed stock markets in a different way than conventional analysis allows. Moreover, phases represented by arrows reveal that the WIG is positively influenced by the PX; these markets also have the largest period of comovement through time and scales.  The PX is also positively correlated with the BUX at several large time and scale periods but the phases do not point to any directional influence. As to the dependence of these markets on the DAX, pair the PX-DAX shows the largest periods of comovement. The WIG is dependent on the DAX while the BUX again shows the weakest dependence through different time and scales periods.

\subsection{Contagion}
The analysis of contagion focuses on the wavelet correlation difference before and after the bankruptcy of the Lehman Brothers in September 2008. The three examined pairs consist of the DAX index and the three CEE indices. We estimate the wavelet correlation on two different time windows, the first window contains observations starting from January 2, 2008 and ending September 15, 2008, the second window begins September 16, 2008 and ends June 5, 2009. Both windows have equal size of 12 860 5--minutes high frequency returns to make them statistically comparable.

\begin{figure}[H]
\centering
\includegraphics[scale=0.65]{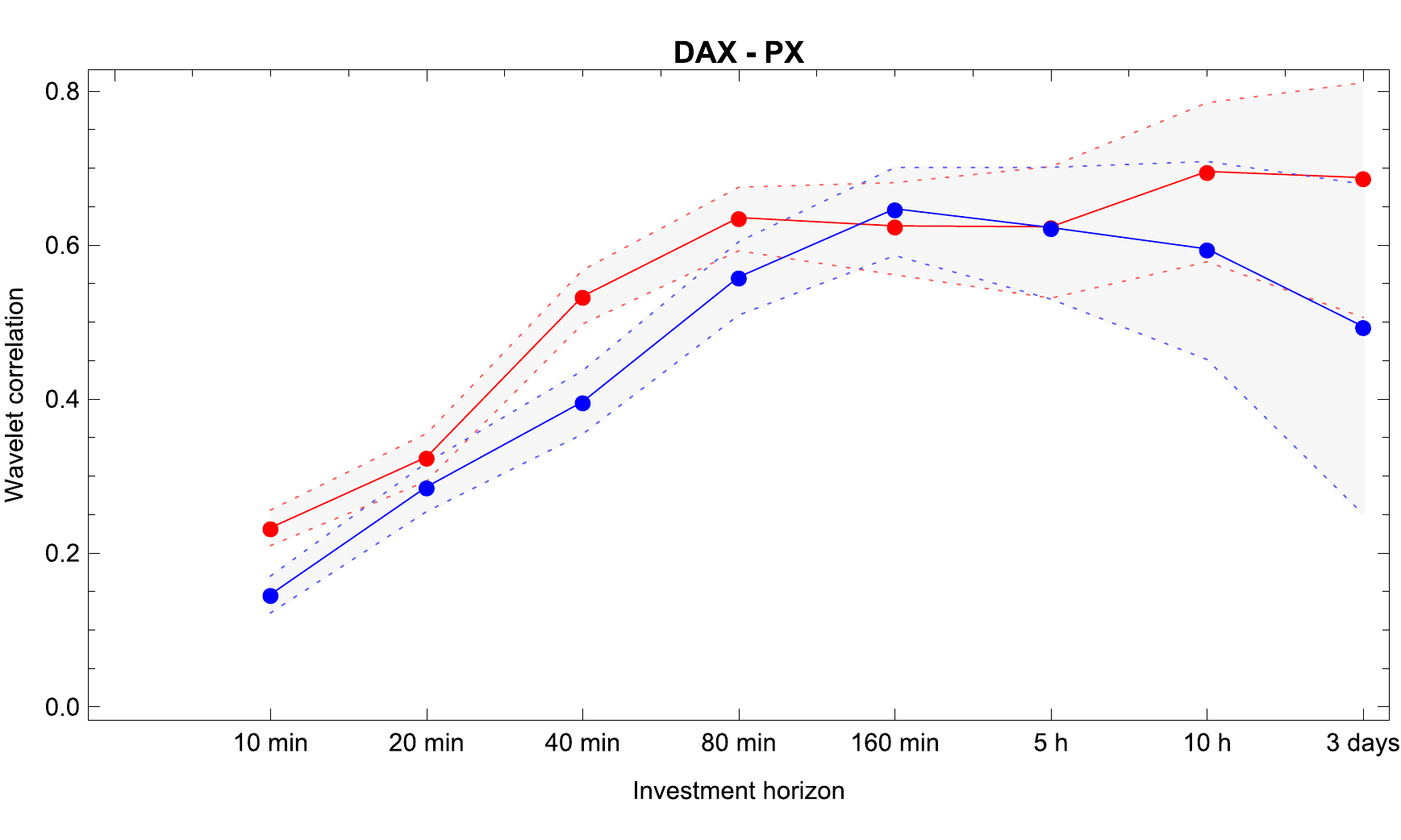}
\caption{Time-frequency correlations of the DAX and the PX. The period before bankruptcy of the Lehman Brothers is depicted by red color, the period after the bankruptcy of the Lehman Brothers by blue color. Grey colored region is the 95\% confidence interval.}
\label{fcontpx}
\end{figure}

\begin{figure}[H]
\centering
\includegraphics[scale=0.65]{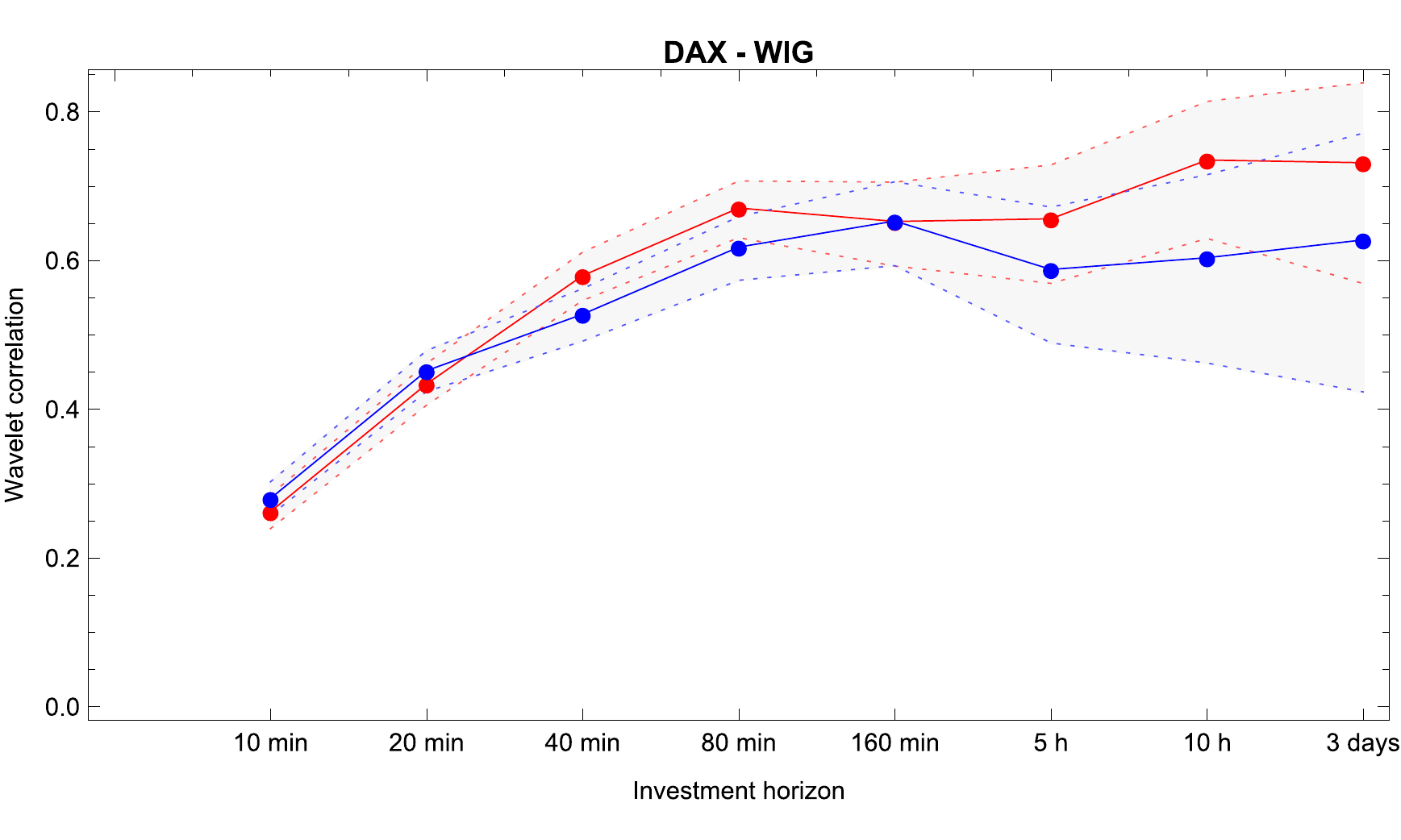}
\caption{Time-frequency correlations of the DAX and the WIG. The period before bankruptcy of the Lehman Brothers is depicted by red color, the period after the bankruptcy of the Lehman Brothers by blue color. Grey colored region is the 95\% confidence interval.}
\label{fcontwig}
\end{figure}

\begin{figure}[H]
\centering
\includegraphics[scale=0.65]{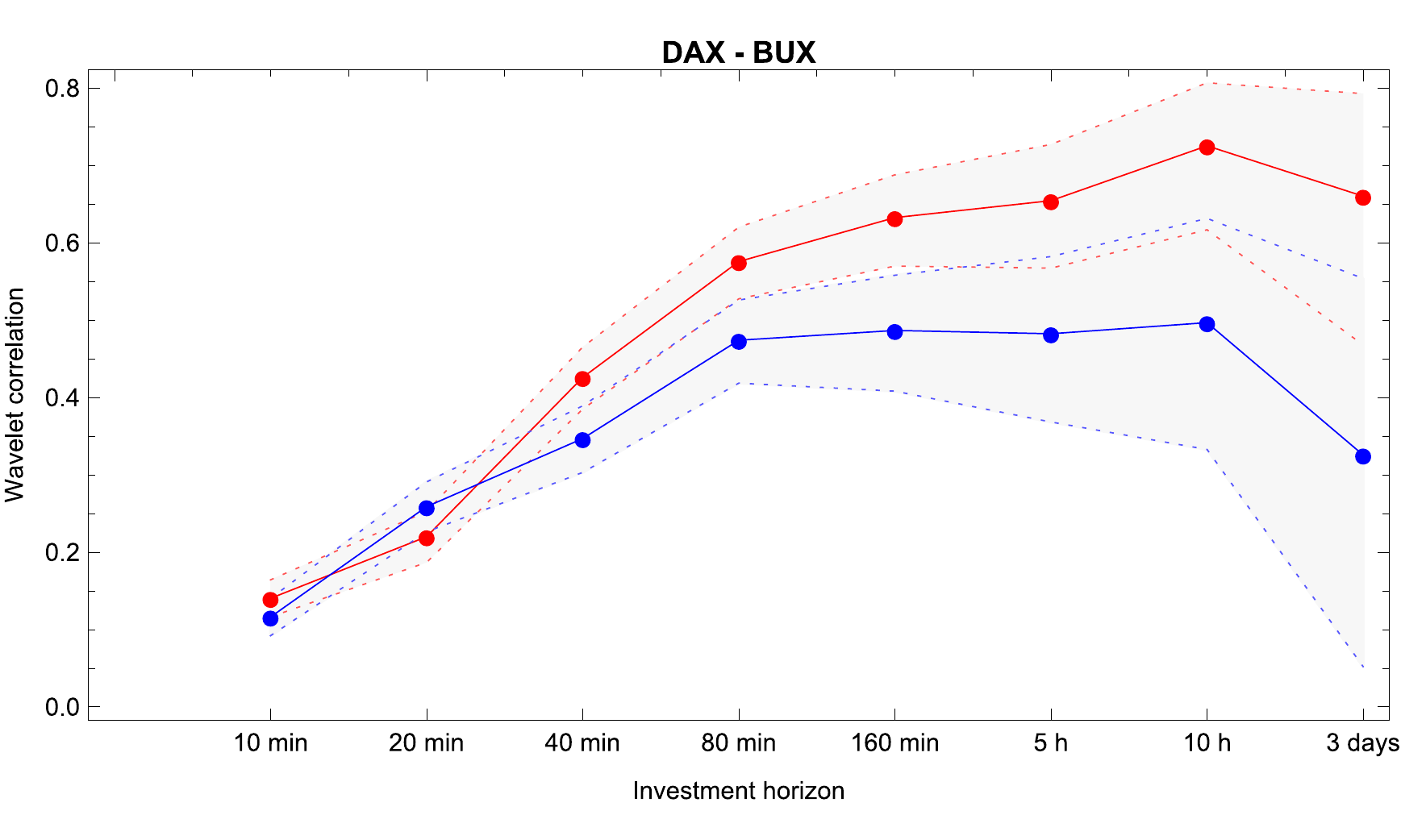}
\caption{Time-frequency correlations of the DAX and the BUX. The period before bankruptcy of the Lehman Brothers is depicted by red color, the period after the bankruptcy of the Lehman Brothers by blue color. Grey colored region is the 95\% confidence interval.}
\label{fcontbux}
\end{figure}

Results of the contagion analysis indicate partial contagion only in case of the PX index, see Figures \ref{fcontpx} -- \ref{fcontbux}. Comparing the wavelet correlation estimates there are only two scales, where the contagion is significant -- scale 1 and 3 -- representing 10 minutes and 80 minutes investment horizons. To sum up the results, we can see that there is only one index out of the three, where we reject the hypothesis of no contagion. However, the result is unexpected because after bankruptcy of the Lehman Brothers, we observe decrease of the wavelet correlations. This result reveals that the comovement between these two markets decreased on short investment horizons. 

Additional interesting aspect arising from the wavelet correlation decomposition is that the CEE markets have low correlations on high frequencies (short investment horizons) with the DAX. This result is in line with findings of \cite{Egert2007}. Our result shows that the CEE markets are not still tightly connected to the leading markets in the region. Interestingly, after the onset of financial crisis this market interconnections lowered even more. While this holds only for the high frequencies, it complements the results from previous literature, which finds interconnections to increase with crisis \citep{gjika2013stock}.

\section{Conclusion}

In this paper, we contribute to the literature on the international stock market comovement and contagion by researching the interconnections between CEE stock markets during the recent crisis in time-frequency space. The novelty of our approach lies in the usage of the wavelet tools to high-frequency financial market data, which allows to understand the relationship between stock market returns in a different way than conventional analysis. Using the wavelet transform, we show how correlations are changing in time and across frequencies, continuously. In the first part of the empirical analysis, we employ the wavelet coherence on high-frequency (5-minutes) data of the Czech (PX), Hungarian (BUX) and Polish (WIG) stock indices with the benchmark German stock index (DAX) in the period of 2008-2009. The second part deals with analysis of contagion in periods before and after the bankruptcy of the Lehman Brothers accompanied with the large crash in September 2008.

The main result of the comovement analysis is finding that interconnection between all stock markets changes significantly in time and varies across frequencies. Using 5--minute high-frequency data, we find the strongest interdependencies among Czech (PX) and Polish (WIG) stock markets. Comovements were significant through various frequencies starting at intraday period and ending at periods up to three months. The PX-WIG pair also shows very interesting development of changing comovements from the second half of year 2008 until the end of the first half of 2009. Correlations are strongly significant through this time period but they change from the one month period (lower frequency) to the shorter period of one week (higher frequency). 

Contagion analysis uncovered partial change in the time-frequency correlation structure for the DAX--PX pair. The first and the third wavelet scale changed significantly after the large 2008 crash, indicating signs of contagion. Unexpectedly, the correlation did not increase, it rather decreased. This result shows that the connection between these two markets decreased on short investment horizons during the crisis. 

Another interesting aspect arising from the correlation decomposition is that the CEE markets generally exhibit low correlation with the DAX on high frequencies. This shows that the CEE markets are still not tightly connected to the leading markets in region. Our results complements the previous literature and opens several interesting venues of research. Time-frequency dynamics can be exploited in terms of forecasting, or risk management. For example, \cite{barunik2008} improves the forecasting ability of the models on CEE data using principal component analysis, hence it would be interesting to see whether decomposition to various investment horizons improve predictive ability.

\section*{Acknowledgements}

We are indebted to Roman Horvath, anonymous referees and seminar participants at the Computational and Financial Econometrics in London (December 2011) for many useful comments, suggestions and discussions. We would like to acknowledge Aslak Grinsted for providing us the MatLab wavelet coherence package. The support of the Czech Science Foundation project No. P402/12/G097 DYME - ``Dynamic Models in Economics" is gratefully acknowledged.

\bibliography{Wavelets}
\bibliographystyle{chicago}

\end{document}